\documentclass[superscriptaddress,twocolumn,amsmath,amssymb,prd,floatfix,nofootinbib]{revtex4}

\usepackage[dvipdfmx]{graphicx}
\usepackage{dcolumn}
\usepackage[varg]{txfonts}
\usepackage{bm}
\usepackage{multirow}

\allowdisplaybreaks

\usepackage{color}

\newcommand*\Laplace{\mathop{}\!\mathbin\bigtriangleup}

\begin{document}

\title{Minkowski functionals and the nonlinear perturbation theory in
  the large-scale structure: second-order effects}

\author{Takahiko Matsubara} \email{tmats@post.kek.jp}
\affiliation{%
  Institute of Particle and Nuclear Studies, High Energy
  Accelerator Research Organization (KEK), Oho 1-1, Tsukuba 305-0801,
  Japan}%
\affiliation{%
  The Graduate University for Advanced Studies (SOKENDAI),
  Tsukuba, Ibaraki 305-0801, Japan}%

\author{Chiaki Hikage} \email{chiaki.hikage@ipmu.jp}
\affiliation{%
  Kavli Institute for the Physics and Mathematics of the Universe
  (Kavli IPMU, WPI), University of Tokyo, 5-1-5 Kashiwanoha, Kashiwa,
  Chiba, 277-8583, Japan}%

\author{Satoshi Kuriki} \email{kuriki@ism.ac.jp}
\affiliation{%
  Institute of Statistical Mathematics, Research Organization of
  Information and Systems, 10-3 Midoricho, Tachikawa, Tokyo 190-8562,
  Japan}%

\date{\today}

\begin{abstract}
    The second-order formula of Minkowski functionals in weakly
    non-Gaussian fields is compared with the numerical $N$-body
    simulations. Recently, weakly non-Gaussian formula of Minkowski
    functionals is extended to include the second-order effects of
    non-Gaussianity in general dimensions. We apply this formula to
    the three-dimensional density field in the large-scale structure
    of the Universe. The parameters of the second-order formula
    include several kinds of skewness and kurtosis parameters. We
    apply the tree-level nonlinear perturbation theory to estimate
    these parameters. First we compare the theoretical values with
    those of numerical simulations on the basis of parameter values,
    and next we test the performance of the analytic formula combined
    with the perturbation theory. The second-order formula outperforms
    the first-order formula in general. The performance of the
    perturbation theory depends on the smoothing radius applied in
    defining the Minkowski functionals. The quantitative comparisons
    are presented in detail.
\end{abstract}


\maketitle


\section{Introduction
\label{sec:Introduction}}

The large-scale structure of the Universe has rich information for
cosmology. The structure originates from the initial density
fluctuations, which are believed to be generated by the cosmic
inflation in the very early universe \cite{Gut81,Sat81,Lin82,Alb82}.
While many scenarios to achieve the inflation are proposed so far
\cite{Mar14}, we still do not know the true mechanism to generate the
initial density field. The large-scale structure of the Universe also
contains information about the evolution of the Universe. Such
cosmological information is contained in the statistical properties of
the large-scale structure, and therefore it is of great importance to
statistically characterize the observed structures.

The power spectrum (and its Fourier counterpart, correlation function)
is one of the most popular statistics to characterize the large-scale
structure \cite{Pee80}. The statistical properties of a random
Gaussian field are completely characterized once the the power
spectrum is specified. The large-scale structure is nearly Gaussian on
sufficiently large scales, or at sufficiently early time, since the
initial condition of the density fluctuations are nearly Gaussian as
indicated by the cosmic microwave background radiation
\cite{Planck18NG}. However, gravitational evolutions destroy the
Gaussianity of the distribution, and non-Gaussianity comes in on small
scales in late time.

How to effectively characterize the non-Gaussian fields is a
nontrivial problem in cosmology. This problem attracts a lot of
attentions because the power spectrum or correlation function cannot
capture the information about the non-Gaussianity. One of the
straightforward way to characterize the non-Gaussianity is to consider
higher-order generalizations of the power spectrum and correlation
function, i.e., polyspectra and $N$-point correlation functions.
Nevertheless, these higher-order correlations are difficult to
accurately measure, because these are functions of scales with many
arguments \cite{Pee80}. There are many alternative methods to
characterize the non-Gaussianity in general.

Among others, the set of Minkowski functionals \cite{Min03,Sch93} is
one of the popular methods to investigate the non-Gaussianity in
cosmology \cite{MBW94,SB97}. Applications of the Minkowski functionals
to the large-scale structure of the Universe are also quite popular
\cite{Ker97,Ker98,Sah98,Sch99,Ker01,She03,Hik03,Sha04,Hik06,Ein11,Liu20}).
The Minkowski functionals are calculated for the excursion set of the
random fields, such as isodensity surfaces of the large-scale
structure. The isodensity surfaces are defined by specifying the
density threshold, and the Minkowski functionals are considered as
functions of the threshold for a given density field.

One of the striking properties of the Minkowski functionals is the
fact that the functional forms of the Minkowski functionals as
functions of the threshold have universal forms for Gaussian random
fields: according to the Tomita's formula \cite{Tom86}, the Minkowski
functionals of random Gaussian fields are represented by specific
functions which are common to all random Gaussian fields. Only the
amplitudes of the functions are affected by the power spectrum of the
distributions. Thus, any deviation from the Gaussian predictions of
the Minkowski functionals as functions of the threshold indicates
non-Gaussianity of the distribution.

Interpreting the deviations from the Gaussian predictions of the
Minkowski functionals is theoretically important to understand the
nature of non-Gaussianity. The theoretical models for the generation
mechanisms of initial density fluctuations usually predict the
higher-order polyspectra such as the bispectrum, trispectrum, and so
forth. The relation between the non-Gaussian Minkowski functionals and
higher-order polyspectra are analytically derived with an expansion
scheme when the non-Gaussianity is weak \cite{TM94,TM03}. The
first-order corrections of the non-Gaussianity in the Minkowski
functionals are solely determined by integrals of the bispectrum,
which are called skewness parameters. Until recently, the analytic
formula of the first-order corrections to the Minkowski functionals
are derived in three or less dimensions. The second-order corrections
depend both on bispectrum and trispectrum. The analytic formula with
second-order corrections in two dimensions is derived \cite{TM10}.
Formal expression of Euler characteristic, or genus statistics, which
is one of the Minkowski functionals, in two and three dimensions in
terms of the Gram-Charlier expansion to all orders are known
\cite{PGP09,GPP12}. Most recently, an analytic formula for
non-Gaussian corrections up to the second order are derived in general
dimensions \cite{KM20,MK20}. The second-order terms involve integrals
of trispectrum, which are called kurtosis parameters. Concrete
relations of the second-order corrections to the bispectrum and
trispectrum are derived in the last literature.

In this paper, we address how the second-order formula works in the
analysis of the large-scale structure in three dimensions. For this
purpose, we employ both the nonlinear perturbation theory and $N$-body
simulations of gravitational evolution in the expanding Universe.
Nonlinear perturbation theory is expected to be valid in weakly
nonlinear regime on large scales, while the $N$-body simulations can
probe the fully nonlinear regime at the expense of computational cost.
The comparison between the perturbation theory and numerical
simulations gives an useful insight into the applicability of the
analytic formula to realistic applications in cosmology.

This paper is organized as follows. In Sec.~\ref{sec:Formula}, the
second-order formula of Minkowski functionals are summarized, and many
parameters in the formula are defined. In Sec.~\ref{sec:PTparam},
methods to evaluate skewness and kurtosis parameters by the nonlinear
perturbation theory of gravitational evolution are developed. In
Sec.~\ref{sec:Numerical}, the analytic formula with the perturbation
theory and the results of $N$-body simulations are compared in detail.
Conclusions are given in Sec.~\ref{sec:Conclusions}.

\section{Analytic formula of Minkowski functionals with
  second-order non-Gaussianity
\label{sec:Formula}}

In this section, we summarize the second-order formula of Minkowski
functionals with weak non-Gaussianity derived in the previous papers
\cite{MK20,KM20}.

First we review mathematical definitions of
Minkowski functionals in three-dimensional density fields
$\rho(\bm{x})$ below \cite{SB97}. We denote the density contrast by
$\delta(\bm{x}) = \rho(\bm{x})/\bar{\rho} - 1$ where
$\bar{\rho} = \langle\rho(\bm{x})\rangle$ is the mean density. In
cosmological applications, the Minkowski functionals are defined in
smoothed density fields,
\begin{equation}
  \label{eq:2-0-1}
  {\delta_\mathrm{s}}(\bm{x}) = \int d^3x' W_R(|\bm{x}-\bm{x}'|)\,
  \delta(\bm{x}'),
\end{equation}
where $W_R(x)$ is a smoothing kernel with smoothing radius $R$. It
is a common practice to apply a Gaussian kernel,
\begin{equation}
  \label{eq:2-0-2}
  W_R(x) = \frac{e^{-x^2/(2R^2)}}{(2\pi)^{3/2}R^3},
\end{equation}
to obtain the smoothed density field. We also assume this kernel
function throughout this paper.

The Minkowski functionals are defined by specifying the isodensity
contours with ${\delta_\mathrm{s}} = \nu\sigma_0$, where $\nu$ is the
threshold and
\begin{equation}
  \label{eq:2-0-3}
  \sigma_0 = \langle{{\delta_\mathrm{s}}}^2\rangle^{1/2}
\end{equation}
is the root-mean-square of the density fluctuations. There are four
Minkowski functionals in three-dimensional space. We denote the
Minkowski functionals per unit volume by $V_k(\nu)$ ($k=0,1,2,3$) as
functions of the threshold $\nu$ which specifies the isodensity
surfaces as defined below.

The Minkowski functional of $k=0$ corresponds to the volume fraction
of the excursion set,
\begin{equation}
  \label{eq:2-1}
  V_0(\nu) = \frac{1}{V} \int_{F_\nu} d^3x,
\end{equation}
where $V$ is the total volume of the sample, and $F_\nu$ is a set of
all positions which satisfies ${\delta_\mathrm{s}} \geq \nu\sigma_0$.
The other Minkowski functionals correspond to surface integrals of the
isodensity surface $\partial F_\nu$, which is the boundary of the
excursion set,
\begin{equation}
  \label{eq:2-2}
  V_k(\nu) = \frac{1}{V} \int_{\partial F_\nu} d^2x\,v_k(\nu,\bm{x}),
\end{equation}
where $v_k(\nu,\bm{x})$ is the local Minkowski functionals defined by
\begin{align}
  \label{eq:2-3a}
  v_1(\nu,\bm{x})
  &= \frac{1}{6},
  \\
  \label{eq:2-3b}
  v_2(\nu,\bm{x})
  &= \frac{1}{6\pi} \left(\frac{1}{R_1} + \frac{1}{R_2}\right),
  \\
  \label{eq:2-3c}
  v_3(\nu,\bm{x})
  &= \frac{1}{4\pi} \frac{1}{R_1 R_2},
\end{align}
and $R_1$, $R_2$ are the radii of curvature of the isodensity surface
orientated toward lower density regions.

The Minkowski functionals have geometrical interpretations: the first
Minkowski functional $V_0$ corresponds to the volume of the excursion
set $F_\nu$ as described above. Minkowski functionals $V_k$ with
$k=1,2$ correspond to the area ($k=1$) and the total mean curvature
($k=2$) of the isodensity surface $\partial F_\nu$, and $V_3$
corresponds the Euler characteristic which is a purely topological
quantity.

Analytic formula of the Minkowski functionals up to second order in
weakly non-Gaussian field in general dimensions $d$ is derived in
Refs.~\cite{MK20,KM20}. In the case of three dimensions, $d=3$, the derived
formula reduces to
\begin{widetext}
\begin{align}
    V_k(\nu)
    &=
    \frac{1}{(2\pi)^{(k+1)/2}}
    \frac{\omega_3}{\omega_{3-k}\omega_k}
    \left(\frac{\sigma_1}{\sqrt{3}\sigma_0}\right)^k
    e^{-\nu^2/2}
    \Biggl[\!\Biggl[
    H_{k-1}(\nu) +
    \left[
    \frac{1}{6}S^{(0)} H_{k+2}(\nu) +
    \frac{k}{3} S^{(1)} H_{k}(\nu)
    +
    \frac{k(k-1)}{6} S^{(2)} H_{k-2}(\nu)
    \right] \sigma_0
    \nonumber\\
  & \qquad
    +
    \Biggl\{
    \frac{1}{72} (S^{(0)})^2 H_{k+5}(\nu)
    +
    \left(
    \frac{1}{24}K^{(0)} + \frac{k}{18} S^{(0)}S^{(1)}
    \right) H_{k+3}(\nu)
    + k
    \left[
    \frac{1}{8} K^{(1)} +
    \frac{k-1}{36} S^{(0)} S^{(2)} +
    \frac{k-2}{18} (S^{(1)})^2
    \right] H_{k+1}(\nu)
    \nonumber\\
  & \hspace{5pc}
    + k
    \left[
    \frac{k-2}{16} K^{(2)}_1 + \frac{k}{16} K^{(2)}_2 +
    \frac{(k-1)(k-4)}{18} S^{(1)}S^{(2)}
    \right] H_{k-1}(\nu)
    \nonumber\\
  & \hspace{8pc}
    +
    k(k-1)(k-2)
    \left[
    \frac{1}{24} K^{(3)} + \frac{k-7}{72} (S^{(2)})^2
    \right] H_{k-3}(\nu)
    \Biggr\} {\sigma_0}^2
    + \mathcal{O}\left({\sigma_0}^3\right)
    \Biggr]\!\Biggr],
    \label{eq:2-3}
\end{align}
\end{widetext}
where $H_n(\nu) = e^{\nu^2/2}(-d/d\nu)^n e^{-\nu^2/2}$ are the
probabilists' Hermite polynomials, and various parameters are given
below in order. First, the factor
\begin{equation}
  \label{eq:2-4}
  \omega_k \equiv \frac{\pi^{k/2}}{\Gamma(k/2+1)}
\end{equation}
is the volume of the unit ball in $k$ dimensions. Second,
\begin{equation}
  \label{eq:2-5}
  \sigma_1 \equiv
  \left\langle \bm{\nabla}{\delta_\mathrm{s}}\cdot\bm{\nabla}{\delta_\mathrm{s}} \right\rangle^{1/2}
\end{equation}
is a spectral moment. Third, $S^{(a)}$ are skewness parameters defined
by
\begin{align}
  &
  S^{(0)} =
  \frac{\left\langle {\delta_\mathrm{s}}^3 \right\rangle_\mathrm{c}}{{\sigma_0}^4},
  \quad
  S^{(1)} = \frac{3}{2}\cdot
  \frac{\left\langle {\delta_\mathrm{s}} |\bm{\nabla} {\delta_\mathrm{s}}|^2\right\rangle_\mathrm{c}}
    {{\sigma_0}^2{\sigma_1}^2},
    \nonumber \\
  &
  S^{(2)}_1 = 
  -\frac{9}{4}\cdot
  \frac{\left\langle |\bm{\nabla} {\delta_\mathrm{s}}|^2 \Laplace
    {\delta_\mathrm{s}}\right\rangle_\mathrm{c}} 
  {{\sigma_1}^4},
  \label{eq:2-6}
\end{align}
where $\langle\cdots\rangle_\mathrm{c}$ denotes the cumulants.
However, all the third-order cumulants in the above equations can
be replaced by simple means, because
$\langle{\delta_\mathrm{s}}\rangle=\langle\bm{\nabla}{\delta_\mathrm{s}}\rangle=0$. Fourth,
$K^{(a)}_\cdot$ are kurtosis parameters defined by
\begin{align}
  \label{eq:2-7a}
  K^{(0)}
  &=
  \frac{\left\langle {\delta_\mathrm{s}}^4 \right\rangle_\mathrm{c}}{{\sigma_0}^6},
  \quad
  K^{(1)} = 2\cdot
  \frac{\left\langle {\delta_\mathrm{s}}^2 |\bm{\nabla} {\delta_\mathrm{s}}|^2\right\rangle_\mathrm{c}}
  {{\sigma_0}^4{\sigma_1}^2},
  \\
  \label{eq:2-7b}
  K^{(2)}_1
  &=
  -\frac{3}{5}\cdot
  \frac{
    5\left\langle {\delta_\mathrm{s}} |\bm{\nabla}{\delta_\mathrm{s}}|^2 \Laplace
    {\delta_\mathrm{s}}\right\rangle_\mathrm{c} +
    \left\langle |\bm{\nabla}{\delta_\mathrm{s}}|^4 \right\rangle_\mathrm{c}
    }
  {{\sigma_0}^2{\sigma_1}^4},
  \\
  \label{eq:2-7c}
  K^{(2)}_2
  &=
  -\frac{3}{5}\cdot
  \frac{5
    \left\langle
      {\delta_\mathrm{s}} |\bm{\nabla}{\delta_\mathrm{s}}|^2 \Laplace {\delta_\mathrm{s}}
    \right\rangle_\mathrm{c} +
    3 \left\langle |\bm{\nabla}{\delta_\mathrm{s}}|^4 \right\rangle_\mathrm{c}} 
  {{\sigma_0}^2{\sigma_1}^4},
  \\
  \label{eq:2-7d}
  K^{(3)}
  &=
  9\cdot
  \frac{
    \left\langle
    |\bm{\nabla}{\delta_\mathrm{s}}|^2 (\Laplace {\delta_\mathrm{s}})^2
    \right\rangle_\mathrm{c} -
    \left\langle
    |\bm{\nabla}{\delta_\mathrm{s}}|^2 \delta_{\mathrm{s},ij}\delta_{\mathrm{s},ij}
    \right\rangle_\mathrm{c}} 
  {{\sigma_1}^6},
\end{align}
where $\delta_{\mathrm{s},ij} \equiv \partial^2{\delta_\mathrm{s}}/\partial x_i\partial x_j$.
The fourth-order cumulants are related to the mean values by
\begin{align}
  \label{eq:2-8a}
  \left\langle {\delta_\mathrm{s}}^4 \right\rangle_\mathrm{c}
  &=
  \left\langle {\delta_\mathrm{s}}^4 \right\rangle - 3 {\sigma_0}^4,
  \\
  \label{eq:2-8b}
  \left\langle {\delta_\mathrm{s}}^2 |\bm{\nabla} {\delta_\mathrm{s}}|^2
  \right\rangle_\mathrm{c}
  &=
  \left\langle {\delta_\mathrm{s}}^2 |\bm{\nabla} {\delta_\mathrm{s}}|^2 \right\rangle
  - {\sigma_0}^2{\sigma_1}^2,
  \\
  \label{eq:2-8c}
  \left\langle
  {\delta_\mathrm{s}} |\bm{\nabla} {\delta_\mathrm{s}}|^2 \triangle {\delta_\mathrm{s}}
  \right\rangle_\mathrm{c}
  &=
  \left\langle {\delta_\mathrm{s}} |\bm{\nabla} {\delta_\mathrm{s}}|^2 \triangle {\delta_\mathrm{s}} \right\rangle
  + {\sigma_1}^4,
  \\
  \label{eq:2-8d}
  \left\langle |\bm{\nabla} {\delta_\mathrm{s}}|^4
  \right\rangle_\mathrm{c}
  &=
  \left\langle |\bm{\nabla} {\delta_\mathrm{s}}|^4 \right\rangle
  - \frac{5}{3} {\sigma_1}^4,
  \\
  \label{eq:2-8e}
  \left\langle |\bm{\nabla} {\delta_\mathrm{s}}|^2
  (\triangle {\delta_\mathrm{s}})^2 \right\rangle_\mathrm{c}
  &=
  \left\langle |\bm{\nabla} {\delta_\mathrm{s}}|^2 (\triangle
    {\delta_\mathrm{s}})^2 \right\rangle 
  - {\sigma_1}^2{\sigma_2}^2,
  \\
  \label{eq:2-8f}
  \left\langle
  |\bm{\nabla} {\delta_\mathrm{s}}|^2 \delta_{\mathrm{s},ij} \delta_{\mathrm{s},ij}
  \right\rangle_\mathrm{c}
  &=
  \left\langle |\bm{\nabla} {\delta_\mathrm{s}}|^2
                             \delta_{\mathrm{s},ij}
                             \delta_{\mathrm{s},ij} \right\rangle 
  - {\sigma_1}^2{\sigma_2}^2,
\end{align}
where 
\begin{equation}
  \label{eq:2-9}
  \sigma_2 \equiv
  \left\langle (\Laplace{\delta_\mathrm{s}})^2 \right\rangle^{1/2}
\end{equation}
is another spectral moment. The formula of Eq.~(\ref{eq:2-3}) is a
generalization of the analytic formula previously derived in
restricted cases \cite{Tom86,TM94,TM03,TM10}.

Various parameters in the formula of Eq.~(\ref{eq:2-3}) are related to
the power spectrum $P(k)$, bispectrum $B(\bm{k}_1,\bm{k}_2,\bm{k}_3)$,
and trispectrum $T(\bm{k}_1,\bm{k}_2,\bm{k}_3,\bm{k}_4)$ of the
(unsmoothed) density contrast $\delta$, which are defined by
\begin{align}
  \label{eq:2-9a}
  &
  \left\langle
  \tilde{\delta}(\bm{k}) \tilde{\delta}(\bm{k}')
  \right\rangle_\mathrm{c}
  = (2\pi)^3 \delta^3(\bm{k}+\bm{k}') P(k),
  \\
  \label{eq:2-9b}
  &
  \left\langle
  \tilde{\delta}(\bm{k}_1) \tilde{\delta}(\bm{k}_2) \tilde{\delta}(\bm{k}_3)
  \right\rangle_\mathrm{c}
  = (2\pi)^3 \delta^3(\bm{k}_1+\bm{k}_2+\bm{k}_3)
    B(\bm{k}_1,\bm{k}_2,\bm{k}_3),
  \\
  &
  \left\langle
  \tilde{\delta}(\bm{k}_1) \tilde{\delta}(\bm{k}_2) \tilde{\delta}(\bm{k}_3)
  \tilde{\delta}(\bm{k}_4) 
    \right\rangle_\mathrm{c}
    \nonumber\\
  \label{eq:2-9c}
  & \qquad
  = (2\pi)^3 \delta^3(\bm{k}_1+\bm{k}_2+\bm{k}_3+\bm{k}_4)
    T(\bm{k}_1,\bm{k}_2,\bm{k}_3,\bm{k}_4),
\end{align}
where
\begin{equation}
    \label{eq:2-10}
    \tilde{\delta}(\bm{k}) = \int d^3\!x\,e^{-i\bm{k}\cdot\bm{x}}
    \delta(\bm{x})
\end{equation}
is the Fourier transform of the density contrast. The Fourier
transform of the smoothed density contrast $\delta_\mathrm{s}$ is
given by
$\tilde{\delta}_\mathrm{s}(\bm{k}) = \tilde{\delta}(\bm{k})W(kR)$,
where
\begin{equation}
  \label{eq:2-10-1}
  W(kR) = \int d^3\!x e^{-i\bm{k}\cdot\bm{x}} W_R(x)
\end{equation}
is a (three-dimensional) Fourier transform of the smoothing kernel. In
the case of Gaussian smoothing, Eq.~(\ref{eq:2-0-2}), we have
\begin{equation}
  \label{eq:2-10-1-1}
  W(kR) = e^{-k^2R^2/2}.
\end{equation}
The smoothed density contrast is therefore given by
\begin{equation}
  \label{eq:2-10-2}
  \delta_\mathrm{s}(\bm{x}) =
  \int \frac{d^3\!k}{(2\pi)^3} e^{i\bm{k}\cdot\bm{x}}
  \tilde{\delta}(\bm{k}) W(kR).
\end{equation}
Substituting Eq.~(\ref{eq:2-10-2}) into Eqs.~(\ref{eq:2-0-3}),
(\ref{eq:2-5})--(\ref{eq:2-7d}), the spectral representations of the
parameters are given by
\begin{align}
  \label{eq:2-11a}
  {\sigma_j}^2
  &= \int \frac{d^3k}{(2\pi)^3} k^{2j} P(k) W^2(kR),
  \\
  \label{eq:2-11b}
  S^{(a)}
  &= \frac{1}{{\sigma_0}^{4-2a}{\sigma_1}^{2a}}
    \int \frac{d^3k_1}{(2\pi)^3} \frac{d^3k_2}{(2\pi)^3}
    \frac{d^3k_3}{(2\pi)^3}
    \nonumber\\
  & \hspace{4pc} \times
    (2\pi)^3\delta^3(\bm{k}_1+\bm{k}_2+\bm{k}_3)
    s^{(a)}(\bm{k}_1,\bm{k}_2,\bm{k}_3)
    \nonumber\\
  & \hspace{4pc} \times
    B(\bm{k}_1,\bm{k}_2,\bm{k}_3)
    W(k_1R) W(k_2R) W(k_3R),
  \\
  K^{(a)}_\cdot
  &= \frac{1}{{\sigma_0}^{6-2a}{\sigma_1}^{2a}}
    \int \frac{d^3k_1}{(2\pi)^3} \frac{d^3k_2}{(2\pi)^3}
    \frac{d^3k_3}{(2\pi)^3}  \frac{d^3k_4}{(2\pi)^3} 
    \nonumber\\
  & \hspace{2pc} \times
    (2\pi)^3\delta^3(\bm{k}_1+\bm{k}_2+\bm{k}_3+\bm{k}_4)
    \kappa^{(a)}_\cdot(\bm{k}_1,\bm{k}_2,\bm{k}_3,\bm{k}_4)
    \nonumber\\
  \label{eq:2-11c}
  & \hspace{2pc} \times
    T(\bm{k}_1,\bm{k}_2,\bm{k}_3,\bm{k}_4)
    W(k_1R) W(k_2R) W(k_3R) W(k_4R),
\end{align}
where
\begin{align}
  &
    s^{(0)} = 1, \quad
    s^{(1)} = -\frac{3}{2} \bm{k}_1\cdot\bm{k}_2, \quad
    s^{(2)} = -\frac{9}{4} (\bm{k}_1\cdot\bm{k}_2){k_3}^2,
    \nonumber \\
  &
    \kappa^{(0)} = 1, \quad
    \kappa^{(1)} = -2\bm{k}_1\cdot\bm{k}_2,
    \nonumber \\
  &
    \kappa^{(2)}_1 =
    -\frac{3}{5}
    (\bm{k}_1\cdot\bm{k}_2)
    \left(
    5{k_3}^2 + \bm{k}_3\cdot\bm{k}_4
    \right),
  \nonumber\\
  &
    \kappa^{(2)}_2 =
    -\frac{3}{5}(\bm{k}_1\cdot\bm{k}_2)
    \left(
    5{k_3}^2 + 3\bm{k}_3\cdot\bm{k}_4
    \right),
  \nonumber\\
  \label{eq:2-12}
  &
    \kappa^{(3)} =
    -9(\bm{k}_1\cdot\bm{k}_2)
    \left[
    {k_3}^2{k_4}^2 - (\bm{k}_3\cdot\bm{k}_4)^2
    \right].
\end{align}
Thus, all the necessary parameters in the formula of
Eq.~(\ref{eq:2-3}) are calculated once the power spectrum, bispectrum,
and trispectrum of the density field is specified.

\section{Evaluating parameters by the nonlinear
  perturbation theory
  \label{sec:PTparam}}

The cosmological perturbation theory of nonlinear density field
\cite{Ber02} is one of the standard methods of evaluating the power
spectrum and higher-order polyspectra in general. Therefore, it is
natural to apply the perturbation theory to predict the parameters of
the formula of non-Gaussian Minkowski functionals. In this section we
derive necessary equations to achieve the evaluations.

\subsection{Spectra from the standard perturbation
  theory
  \label{subsec:SPT}}

In the standard perturbation theory, the nonlinear density contrast
$\tilde{\delta}(\bm{k})$ in Fourier space is expanded by the linear
density contrast $\delta_\mathrm{L}(\bm{k})$ as
\begin{multline}
  \label{eq:3-1}
  \tilde{\delta}(\bm{k}) =
  \sum_{n=1}^\infty \frac{1}{n!}
  \int
  \frac{d^3k_1}{(2\pi)^3} \cdots \frac{d^3k_n}{(2\pi)^3}
  \delta_\mathrm{D}^3(\bm{k}_1+\cdots +\bm{k}_n-\bm{k})
  \\
  \times
  F_n(\bm{k}_1,\ldots,\bm{k}_n)
  \delta_\mathrm{L}(\bm{k}_1) \cdots \delta_\mathrm{L}(\bm{k}_n),
\end{multline}
and the similar expansion is applied to the velocity (divergence)
field $\theta$ with kernel functions
$G_n(\bm{k}_1,\ldots,\bm{k}_n)$\footnote{Our conventions for the
  kernel functions $F_n$ and $G_n$ are different from most of the
  literatures in which a factor $1/n!$ in Eq.~(\ref{eq:3-1}) is
  missing. One should replace $F_n, G_n \rightarrow n!F_n, n!G_n$ to
  reproduce the equations in the corresponding literatures. Our
  conventions designate most of the derived equations more concise.}.
Using the recursion relations \cite{Gor86,Ber02} of the kernels $F_n$
and $G_n$, we have
\begin{align}
  \label{eq:3-2a}
  F_2(\bm{k}_1,\bm{k}_2) &=
  \frac{10}{7} +
  \left(\frac{k_1}{k_2}+\frac{k_2}{k_1}\right)
  \frac{\bm{k}_1\cdot\bm{k}_2}{k_1k_2} +
  \frac{4}{7}
  \left(
  \frac{\bm{k}_1\cdot\bm{k}_2}{k_1k_2}
  \right)^2,
  \\
  \label{eq:3-2b}
  G_2(\bm{k}_1,\bm{k}_2) &=
  \frac{6}{7} +
  \left(\frac{k_1}{k_2}+\frac{k_2}{k_1}\right)
  \frac{\bm{k}_1\cdot\bm{k}_2}{k_1k_2} +
  \frac{8}{7}
  \left(
  \frac{\bm{k}_1\cdot\bm{k}_2}{k_1k_2}
  \right)^2,
\end{align}
and
\begin{multline}
  \label{eq:3-3}
  F_3(\bm{k}_1,\bm{k}_2,\bm{k}_3) =
  \frac{7}{6} \alpha(\bm{k}_1,\bm{k}_2+\bm{k}_3)
  F_2(\bm{k}_2,\bm{k}_3)
  \\
  + \frac{1}{6}
  \left[
    7\alpha(\bm{k}_2+\bm{k}_3,\bm{k}_1) +
    4\beta(\bm{k}_2+\bm{k}_3,\bm{k}_1)
  \right] G_2(\bm{k}_2,\bm{k}_3),
\end{multline}
where
\begin{equation}
  \label{eq:3-4}
  \alpha(\bm{k}_1,\bm{k}_2) \equiv
  1 + \frac{\bm{k}_1\cdot\bm{k}_2}{{k_1}^2}, \quad
  \beta(\bm{k}_1,\bm{k}_2) \equiv
  \frac{|\bm{k}_1+\bm{k}_2|^2(\bm{k}_1\cdot\bm{k}_2)}
  {2{k_1}^2{k_2}^2}.
\end{equation}
Instead of the asymmetric kernel $F_3$ in the above equation, it is
convenient to define the symmetrized kernel
\begin{equation}
  \label{eq:3-5}
  F^\mathrm{(s)}_3(\bm{k}_1,\bm{k}_2,\bm{k}_3)
  \equiv
  \frac{1}{3}F_3(\bm{k}_1,\bm{k}_2,\bm{k}_3) + \mathrm{cyc.},
\end{equation}
where ``$+\,\mathrm{cyc.}$'' denotes the cyclic permutations of the
previous term.

In the lowest-order approximations (so-called ``tree-level''
approximations) of the perturbation theory, the power spectrum,
bispectrum and trispectrum defined in
Eqs.~(\ref{eq:2-9a})--(\ref{eq:2-9c}) are given by
\begin{align}
  \label{eq:3-6a}
  P(k)
  &= P_\mathrm{L}(k),
  \\
  \label{eq:3-6b}
  B(\bm{k}_1,\bm{k}_2,\bm{k}_3)
  &=
  F_2(\bm{k}_1,\bm{k}_2) P_\mathrm{L}(k_1) P_\mathrm{L}(k_2) +
  \mathrm{cyc.},
  \\
  \label{eq:3-6c}
  T(\bm{k}_1,\bm{k}_2,\bm{k}_3,\bm{k}_4)
  &=
    F_2(\bm{k}_1,\bm{k}_2+\bm{k}_3) F_2(\bm{k}_2,-\bm{k}_2-\bm{k}_3)
    \nonumber\\
  &\quad \times
    P_\mathrm{L}(k_1) P_\mathrm{L}(k_2)
    P_\mathrm{L}(|\bm{k}_2+\bm{k}_3|) +
    \mathrm{sym.(11)}
    \nonumber\\
  &\quad +
    F^\mathrm{(s)}_3(\bm{k}_1,\bm{k}_2,\bm{k}_3)
    P_\mathrm{L}(k_1) P_\mathrm{L}(k_2) P_\mathrm{L}(k_3)
    \nonumber\\
  &\qquad +
    \mathrm{sym.}(3),
\end{align}
where ``$+\,\mathrm{sym.}(n)$'' represents additional $n$ terms to
symmetrize the previous term with respect to the arguments
$\bm{k}_1,\ldots,\bm{k}_4$. Substituting
Eqs.~(\ref{eq:3-6a})--(\ref{eq:3-6c}) and (\ref{eq:2-12}) into
Eqs.~(\ref{eq:2-11a})--(\ref{eq:2-11c}), the parameters of the
second-order formula (\ref{eq:2-3}) of Minkowski functionals are given
in the tree-level perturbation theory of the gravitational evolution
of density field. However, it is not straightforward to numerically
evaluate the skewness and kurtosis parameters with above equations as
they involve higher-dimensional integrals. One can analytically reduce
the dimensionality of the integrals in order to practically evaluate
them as we explain next.

\subsection{Skewness and kurtosis parameters
  \label{subsec:SKparam}}

The method to evaluate the skewness parameters $S^{(a)}$ with the
Gaussian smoothing kernel in the perturbation theory are already known
\cite{TM94}. The simplest kurtosis $K^{(0)}$ with the Gaussian
smoothing kernel are also already addressed \cite{Lok95}. We follow a
similar, but somehow different approach to achieve the numerical
evaluations of all the parameters. For that purpose, it turns out to
be desirable to reexpress the integrals of skewness and kurtosis
parameters, Eqs.~(\ref{eq:2-11b}) and
(\ref{eq:2-11c}). The functions $s^{(a)}$ and $\kappa^{(a)}_\cdot$ can
be replaced by the symmetrized ones,
\begin{align}
  \label{eq:3-10a}
  \tilde{s}^{(a)}(\bm{k}_1,\bm{k}_2,\bm{k}_3)
  &\equiv
  \frac{1}{3} s^{(a)}(\bm{k}_1,\bm{k}_2,\bm{k}_3)
  + \mathrm{cyc.},
  \\
  \label{eq:3-10b}
  \tilde{\kappa}^{(a)}_\cdot(\bm{k}_1,\bm{k}_2,\bm{k}_3,\bm{k}_4)
  &\equiv
  \frac{1}{12}
  \kappa^{(a)}_\cdot(\bm{k}_1,\bm{k}_2,\bm{k}_3,\bm{k}_4)
  + \mathrm{sym.(11)}.
\end{align}
These functions are completely symmetric for any permutations of their
arguments. When one replaces $s^{(a)} \rightarrow \tilde{s}^{(a)}$ and
$\kappa^{(a)}_\cdot \rightarrow \tilde{\kappa}^{(a)}_\cdot$ in
Eqs.~(\ref{eq:2-11b}) and (\ref{eq:2-11c}), the bispectrum and
trispectrum in the perturbation theory can be replaced by asymmetric
functions,
\begin{align}
  \label{eq:3-11a}
  \tilde{B}(\bm{k}_1,\bm{k}_2)
  &\equiv
  3 F_2(\bm{k}_1,\bm{k}_2) P_\mathrm{L}(k_1) P_\mathrm{L}(k_2),
  \\
  \label{eq:3-11b}
  \tilde{T}(\bm{k}_1,\bm{k}_2,\bm{k}_3)
  &\equiv
    12 F_2(\bm{k}_1,\bm{k}_2+\bm{k}_3) F_2(\bm{k}_2,-\bm{k}_2-\bm{k}_3)
    \nonumber\\
  &\qquad \times
    P_\mathrm{L}(k_1) P_\mathrm{L}(k_2)
    P_\mathrm{L}(|\bm{k}_2+\bm{k}_3|)
    \nonumber\\
  &\quad +
    4 F_3(\bm{k}_1,\bm{k}_2,\bm{k}_3)
    P_\mathrm{L}(k_1) P_\mathrm{L}(k_2) P_\mathrm{L}(k_3).
\end{align}

Because of the delta functions in Eqs.~(\ref{eq:2-11b}) and
(\ref{eq:2-11c}), one can replace $\bm{k}_3 = -\bm{k}_1-\bm{k}_2$
in Eq.~(\ref{eq:3-10a}) and $\bm{k}_4 = -\bm{k}_1-\bm{k}_2-\bm{k}_3$
in Eq.~(\ref{eq:3-10b}). We denote
$\tilde{s}^{(a)}(\bm{k}_1,\bm{k}_2)$ and
$\tilde{\kappa}^{(a)}_\cdot(\bm{k}_1,\bm{k}_2,\bm{k}_3)$ after
substituting these constraints. After all, Eqs.~(\ref{eq:2-11b}) and
(\ref{eq:2-11c}) are reexpressed as
\begin{align}
  \label{eq:3-13a}
  S^{(a)}
  &= \frac{1}{{\sigma_0}^{4-2a}{\sigma_1}^{2a}}
    \int \frac{d^3k_1}{(2\pi)^3} \frac{d^3k_2}{(2\pi)^3}
    \tilde{s}^{(a)}(\bm{k}_1,\bm{k}_2)
    \nonumber\\
  & \hspace{6pc} \times
    \tilde{B}(\bm{k}_1,\bm{k}_2)
    e^{-({k_1}^2 + {k_2}^2 + \bm{k}_1\cdot\bm{k}_2)R^2},
  \\
  \label{eq:3-13b}
  K^{(a)}_\cdot
  &= \frac{1}{{\sigma_0}^{6-2a}{\sigma_1}^{2a}}
    \int \frac{d^3k_1}{(2\pi)^3} \frac{d^3k_2}{(2\pi)^3}
    \frac{d^3k_3}{(2\pi)^3}
    \tilde{\kappa}^{(a)}_\cdot(\bm{k}_1,\bm{k}_2,\bm{k}_3)
    \nonumber\\
  & \hspace{5pc} \times
    \tilde{T}(\bm{k}_1,\bm{k}_2,\bm{k}_3)
    \nonumber\\
  & \hspace{5pc} \times
     e^{-({k_1}^2 + {k_2}^2 + {k_3}^2
     + \bm{k}_1\cdot\bm{k}_2 + \bm{k}_2\cdot\bm{k}_3
     + \bm{k}_3\cdot\bm{k}_1)R^2},
\end{align}
where
\begin{align}
  \label{eq:3-14a}
  \tilde{s}^{(0)}
  &= 1,
  \\
  \label{eq:3-14b}
  \tilde{s}^{(1)}
  &= \frac{1}{2}
    \left({k_1}^2 + {k_2}^2 + \bm{k}_1\cdot\bm{k}_2\right),
  \\
  \label{eq:3-14c}
  \tilde{s}^{(2)}
  &= \frac{3}{2}
    \left[
    {k_1}^2 {k_2}^2 - (\bm{k}_1\cdot\bm{k}_2)^2
    \right],
\end{align}
and
\begin{align}
  \label{eq:3-15a}
  \tilde{\kappa}^{(0)}
  &= 1,
  \\
  \label{eq:3-15b}
  \tilde{\kappa}^{(1)}
  &= \frac{1}{3}
    \left(
    {k_1}^2 + {k_2}^2 + {k_3}^2
    + \bm{k}_1\cdot\bm{k}_2
    + \bm{k}_2\cdot\bm{k}_3
    + \bm{k}_3\cdot\bm{k}_1
    \right),
  \\
  \label{eq:3-15c}
  \tilde{\kappa}^{(2)}_1
  &= \frac{1}{10}
    \biggl\{
    5 \left[ {k_1}^2 {k_2}^2 - (\bm{k}_1\cdot\bm{k}_2)^2 \right]
    \nonumber\\
  & \hspace{2.2pc}
    - 6 (\bm{k}_1\cdot\bm{k}_3)(\bm{k}_2\cdot\bm{k}_3)
    + 2 (\bm{k}_1\cdot\bm{k}_2) {k_3}^2
    \biggr\}
    + \mathrm{cyc.},
  \\
  \label{eq:3-15d}
  \tilde{\kappa}^{(2)}_2
  &= \frac{1}{10}
    \biggl\{
    5 \left[ {k_1}^2 {k_2}^2 - (\bm{k}_1\cdot\bm{k}_2)^2 \right]
    \nonumber\\
  & \hspace{2.2pc}
    + 2 (\bm{k}_1\cdot\bm{k}_3)(\bm{k}_2\cdot\bm{k}_3)
    + 6 (\bm{k}_1\cdot\bm{k}_2) {k_3}^2
    \biggr\}
    + \mathrm{cyc.},
  \\
  \label{eq:3-15e}
  \tilde{\kappa}^{(3)}
  &= \frac{3}{2}
    \biggl[
    {k_1}^2 {k_2}^2 {k_2}^3 +
    2 (\bm{k}_1\cdot\bm{k}_2) (\bm{k}_2\cdot\bm{k}_3)
    (\bm{k}_3\cdot\bm{k}_1)
    \nonumber\\
  & \hspace{8pc}
    - 3 (\bm{k}_1\cdot\bm{k}_2)^2 {k_3}^2
    \biggr]
    + \mathrm{cyc.},
\end{align}
and the Gaussian window function, Eq.~(\ref{eq:2-10-1-1}), is
explicitly used. In the lowest-order in the perturbation theory, the
parameters $\sigma_j$ of Eq.~(\ref{eq:2-11a}) are given by
\begin{equation}
  \label{eq:3-15-1}
  {\sigma_j}^2 =
  \int_0^\infty \frac{k^2dk}{2\pi^2}
  k^{2j}P_\mathrm{L}(k)e^{-k^2R^2},
\end{equation}
where the Gaussian window function is assumed.

Because the integrands of Eqs.~(\ref{eq:3-13a}) and (\ref{eq:3-13b})
are rotationally invariant, one can reduce the dimensionality of the
integrals by three dimensions. In order to reduce the dimensionality
of integrals of Eq.~(\ref{eq:3-13a}) for the skewness parameters, one
can choose coordinates system,
\begin{align}
  \label{eq:3-16a}
  \bm{k}_1 R = (p\sin\theta,0,p\cos\theta), \quad
  \bm{k}_2 R = (0,0,q),
\end{align}
and the volume element of the integral in Eq.~(\ref{eq:3-13a}) reduces
to
\begin{equation}
  \label{eq:3-17}
  \int \frac{d^3k_1}{(2\pi)^3} \frac{d^3k_2}{(2\pi)^3}
  \rightarrow
  \frac{1}{8\pi^4R^6} \int_0^\infty\! p^2dp \int_0^\infty\! q^2dq
  \int_0^\pi \!\sin\theta\,d\theta.
\end{equation}
We substitute Eqs.~(\ref{eq:3-11a}) and
(\ref{eq:3-14a})--(\ref{eq:3-14c}) into Eq.~(\ref{eq:3-13a}) in this
coordinate system. The integral over the variable $\theta$ can
analytically performed as
\begin{equation}
  \label{eq:3-18}
  \int_0^\pi \sin\theta\,d\theta\,e^{x\,\cos\theta} \cos^n\theta =
  2\frac{d^n}{dx^n}\left(\frac{\sinh x}{x}\right).
\end{equation}
As a result, we have expressions in a form,
\begin{align}
  \label{eq:3-19}
  S^{(a)}
  &= \frac{1}{8\pi^4R^{2a+6}{\sigma_0}^{4-2a}{\sigma_1}^{2a}}
    \nonumber\\
  & \quad \times
    \int_0^\infty dp\,dq\,
    e^{-p^2-q^2}
    \tilde{S}^{(a)}(p,q)
    P_\mathrm{L}\left(\frac{p}{R}\right)
    P_\mathrm{L}\left(\frac{q}{R}\right),
\end{align}
where $\tilde{S}^{(a)}(p,q)$ are analytic functions which are
explicitly given by
\begin{widetext}
\begin{align}
  \label{eq:3-20a}
  \tilde{S}^{(0)}
  &=
    - 6
    \left(
    p^2 + q^2 + \frac{8}{7}
    \right)
    \cosh (pq)
    + 6
    \left(
    2p^2q^2 + p^2 + q^2 + \frac{8}{7}
    \right)
    \frac{\sinh (pq)}{pq},
  \\
  \label{eq:3-20b}
  \tilde{S}^{(1)}
  &=
    - 3
    \left[
    p^4 + q^4 + 4 p^2 q^2 + \frac{22}{7} (p^2 + q^2)
    + \frac{24}{7}
    \right]
    \cosh (pq)
    + 3
    \left[
    \left( 3p^2q^2  + \frac{22}{7} \right) (p^2 + q^2)
    +\, p^4 + q^4 + \frac{36}{7} p^2 q^2
    + \frac{24}{7}
    \right]
    \frac{\sinh (pq)}{pq},
  \\
  \label{eq:3-20c}
  \tilde{S}^{(2)}
  &=
    18
    \left[
    2 p^2 q^2 + 3(p^2 + q^2)
    + \frac{48}{7}
    \right]
    \cosh (pq)
    -18
    \left[
    \left( p^2q^2  + 3 \right) (p^2 + q^2)
    + \frac{30}{7} p^2 q^2
    + \frac{48}{7}
    \right]
    \frac{\sinh (pq)}{pq}.
\end{align}
\end{widetext}
The two-dimensional integrations of Eq.~(\ref{eq:3-19}) are
numerically evaluated without any difficulty.

Similarly, the dimensionality of integrals of Eq.~(\ref{eq:3-13b}) for
kurtosis parameters can be reduced due to the rotational invariance of
integrands. It is convenient to change the integration variables
\cite{Lok95},
\begin{equation}
  \label{eq:3-21}
  \bm{p} = \bm{k}_1 R, \quad
  \bm{q} = \bm{k}_2 R, \quad
  \bm{r} = (\bm{k}_2+\bm{k}_3) R,
\end{equation}
or,
\begin{equation}
  \label{eq:3-22}
  \bm{k}_1 = \frac{\bm{p}}{R}, \quad
  \bm{k}_2 = \frac{\bm{q}}{R}, \quad
  \bm{k}_3 = \frac{\bm{r}-\bm{q}}{R}.
\end{equation}
One can choose coordinates system,
\begin{align}
  \label{eq:3-23a}
  \bm{p}
  &= (p\sin\theta\cos\phi,p\sin\theta\sin\phi,p\cos\theta)
  \\         
  \label{eq:3-23b}
  \bm{q}
  &= (q\sin\theta',0,q\cos\theta'), \quad
    \bm{r} = (0,0,r).
\end{align}
and the volume element of the integral in Eq.~(\ref{eq:3-13b}) reduces
to
\begin{multline}
  \label{eq:3-24}
  \int \frac{d^3k_1}{(2\pi)^3} \frac{d^3k_2}{(2\pi)^3}
  \frac{d^3k_3}{(2\pi)^3}
  \\
  \rightarrow
  \frac{1}{32\pi^6R^9} \int_0^\infty\! p^2dp \int_0^\infty\! q^2dq
  \int_0^\infty\! r^2dr
  \\
  \times
  \int_0^\pi \!\sin\theta\,d\theta
  \int_0^\pi \!\sin\theta'\,d\theta'
  \int_0^{2\pi}\!\frac{d\phi}{2\pi}.
\end{multline}
The integral over the variable $\phi$ is straightforward, and the
integral over $\theta$ can again analytically performed by applying
Eq.~(\ref{eq:3-18}). The integrals over $\theta'$ are only possible
for the first term of Eq.~(\ref{eq:3-11b}), and are not possible for
the second term. We use a new integration variable $\mu=\cos\theta'$
for the last integrals. As a result, we have expressions in a form,
\begin{align}
  \label{eq:3-25}
  K^{(a)}_\cdot
  &= \frac{1}{32\pi^6R^{2a+9}{\sigma_0}^{6-2a}{\sigma_1}^{2a}}
    \int_0^\infty dp\,dq\,dr\,
    e^{-p^2-q^2-r^2}
    \nonumber\\
  & \qquad \times
    \Biggl[
    \tilde{K}^{(a)}_\cdot(p,q,r)
    P_\mathrm{L}\left(\frac{p}{R}\right)
    P_\mathrm{L}\left(\frac{q}{R}\right)
    P_\mathrm{L}\left(\frac{r}{R}\right)
    \nonumber\\
  & \qquad\qquad +
    \int_{-1}^1d\mu\,e^{qr\mu}
    \tilde{L}^{(a)}_\cdot(p,q,r,\mu)
    P_\mathrm{L}\left(\frac{p}{R}\right)
    P_\mathrm{L}\left(\frac{q}{R}\right)
    \nonumber\\
  & \hspace{7.7pc} \times
    P_\mathrm{L}\left(\frac{\sqrt{q^2 + r^2 - 2qr\mu}}{R}\right)
    \Biggr],
\end{align}
where
$\tilde{K}^{(a)}_\cdot$, $\tilde{L}^{(a)}_\cdot$ are analytic
functions. For example,
\begin{widetext}
\begin{align}
  \label{eq:3-26a}
  \tilde{K}^{(0)}
  &=
    \frac{48}{r^2}
    \left[
    \left(p^2 + r^2 + \frac{8}{7}\right)\cosh(pr)
    - \left(2p^2r^2 + p^2 + r^2 + \frac{8}{7}\right)
    \frac{\sinh(pr)}{pr}
    \right]
    \nonumber\\
  & \hspace{7pc} \times
    \left[
    \left(q^2 + r^2 + \frac{8}{7}\right)\cosh(qr)
    - \left(2q^2r^2 + q^2 + r^2 + \frac{8}{7}\right)
    \frac{\sinh(qr)}{qr}
    \right],
  \\  
  \label{eq:3-26b}
  \tilde{L}^{(0)}
  &= \frac{4qr^2}{21(q^2 + r^2 - 2 qr\mu)}
    \biggl\{
    \left[
    \left(8 + 9p^2 -19r^2\right)q - 7\left(8+9p^2+9r^2\right)r\mu
    + 2\left(24+27p^2+41r^2\right)q\mu^2
    \right]\cosh(pr)
    \nonumber\\
  &\hspace{8pc}
    -
    \left[
    \left(8 +9p^2-19r^2-10p^2r^2\right)q
    - 7\left(8+9p^2+9r^2+18p^2r^2\right)r\mu
   \right.
    \nonumber\\
  &\hspace{19pc}
    \left.
    +\, 2\left(24+27p^2+41r^2+68p^2r^2\right)q\mu^2
    \right]
    \frac{\sinh(pr)}{pr}
    \biggr\}.
\end{align}
\end{widetext}
Other functions $\tilde{K}^{(1)}$, $\tilde{K}^{(2)}_1$,
$\tilde{K}^{(2)}_2$, $\tilde{K}^{(3)}$ are similarly given, although
explicit expressions of these functions are too tedious to reproduce
here. It is straightforward to derive the expressions by using
\textsc{Mathematica} package. With these analytic results, we
numerically evaluate the three- and four-dimensional integrations of
Eq.~(\ref{eq:3-25}).

\section{Comparisons with numerical simulations
  \label{sec:Numerical}}

\begin{table*}
\centering
\caption{\label{tab:params} The values of parameters for the weakly
  non-Gaussian formula of Minkowski functionals in the large-scale
  structure. Four cases of the smoothing radius $R$ are presented. The
  values calculated from the lowest-order perturbation theory (upper
  figures) and directly from the numerical simulations with $1\sigma$
  errors (lower figures) are listed. }
\begin{tabular}{ccccc}
  \hline
  $R$ [$h^{-1}$Mpc]
  & $10$  & $20$  & $30$  & $40$ \\
\hline\hline
  \multirow{2}{*}{$\sigma_0$}
  & $0.385$ & $0.193$ & $0.121$ & $0.0845$ \\
  & $0.3804 \pm 0.0001$ & $0.1899 \pm 0.0001$
  & $0.1194 \pm 0.0001$ & $0.08374 \pm 0.0001$\\ 
  \hline
  \multirow{2}{*}{$\sigma_1$}
  & $0.0367$ & $0.0101$ & $0.00441$ & $0.00240$ \\
  & $0.03652 \pm 0.00001$ & $0.009918 \pm 0.000004$
  & $0.004352 \pm 0.000003$ & $0.002371 \pm 0.000002$\\
  \hline
  \multirow{2}{*}{$S^{(0)}$}
  & $3.56$ & $3.40$ & $3.33$ & $3.28$ \\
  & $3.762 \pm 0.004$ & $3.46 \pm 0.01$ & $3.36 \pm 0.02$
  & $3.28 \pm 0.05$\\
  \hline
  \multirow{2}{*}{$S^{(1)}$}
  & $3.63$ & $3.45$ & $3.36$ & $3.31$ \\
  & $3.932 \pm 0.003$ & $3.531 \pm 0.006$ & $3.41 \pm 0.01$
  & $3.34 \pm 0.03$\\
  \hline
  \multirow{2}{*}{$S^{(2)}$}
  & $3.66$ & $3.68$ & $3.71$ & $3.72$ \\
  & $4.499 \pm 0.004$ & $3.887 \pm 0.006$ & $3.81 \pm 0.01$
  & $3.78 \pm 0.03$\\
  \hline
  \multirow{2}{*}{$K^{(0)}$}
  & $23.2$ & $20.9$ & $19.9$ & $19.2$ \\
  & $26.66 \pm 0.09$ & $21.6 \pm 0.2$ & $20.2 \pm 0.5$
  & $19 \pm 1$\\
  \hline
  \multirow{2}{*}{$K^{(1)}$}
  & $23.8$ & $21.3$ & $20.2$ & $19.5$ \\
  & $28.77 \pm 0.09$ & $22.2 \pm 0.1$ & $20.6 \pm 0.3$
  & $19.8 \pm 0.8$\\
  \hline
  \multirow{2}{*}{$K^{(2)}_1$}
  & $30.6$ & $28.3$ & $27.2$ & $26.7$ \\
  & $41.7 \pm 0.1$ & $30.4 \pm 0.2$ & $28.1 \pm 0.4$
  & $27.5 \pm 0.8$\\
  \hline
  \multirow{2}{*}{$K^{(2)}_2$}
  & $18.8$ & $17.7$ & $17.3$ & $17.0$ \\
  & $26.7 \pm 0.1$ & $19.2 \pm 0.1$ & $18.0 \pm 0.2$
  & $17.8 \pm 0.6$\\
  \hline
  \multirow{2}{*}{$K^{(3)}$}
  & $25.1$ & $25.6$ & $26.1$ & $26.4$ \\
  & $45.5 \pm 0.2$ & $29.6 \pm 0.2$ & $27.6 \pm 0.4$
  & $27 \pm 1$\\
  \hline
\end{tabular}
\end{table*}

To see how the non-Gaussian formula of Minkowski functionals work in
the three-dimensional large-scale structure, we now compare the
analytic predictions and the results of cosmological $N$-body
simulations. For that purpose, we calculate the Minkowski functionals
from 300 realizations of the $N$-body simulations from the
\textsc{Quijote} suite \cite{Quijote}. Each realization contains
$N=512^3$ particles in a box size of $V=1\,h^{-3}\mathrm{Gpc}^3$. The
cosmological parameters of the simulations are given by
$\Omega_\mathrm{m}=0.3175$, $\Omega_\mathrm{b}=0.049$, $h=0.6711$,
$n_\mathrm{s}=0.9624$, $\sigma_8=0.834$, and a flat {$\Lambda$}CDM
model is assumed. In calculating the Minkowski functionals, the
density field in the simulation box is smoothed by a Gaussian filter
of the radius $R=10,20,30,40\,h^{-1}\mathrm{Mpc}$. The Minkowski
functionals are numerically evaluated based on Crofton's formula from
integral geometry \cite{Had57,Cro1868} where each Minkowski functional
can be computed by counting the numbers of vertices, edges, faces, and
cubes of the excursion set over a threshold $\nu$ \cite{SB97}.

With the same set of cosmological parameters, theoretical predictions
of the weakly non-Gaussian formula of Minkowski functionals with the
nonlinear perturbation theory are calculated according to the method
described in the previous section. The linear power spectrum
$P_\mathrm{L}(k)$ is evaluated by the \textsc{CLASS} code
\cite{class11,CLASS}. Once the linear power spectrum is given, all the
parameters in the formula of Eq.~(\ref{eq:2-3}) for the Minkowski
functionals are calculated by numerical integrations of
Eqs.~(\ref{eq:3-15-1}), (\ref{eq:3-19}), (\ref{eq:3-25}).

\begin{figure}
  \includegraphics[width=20pc]{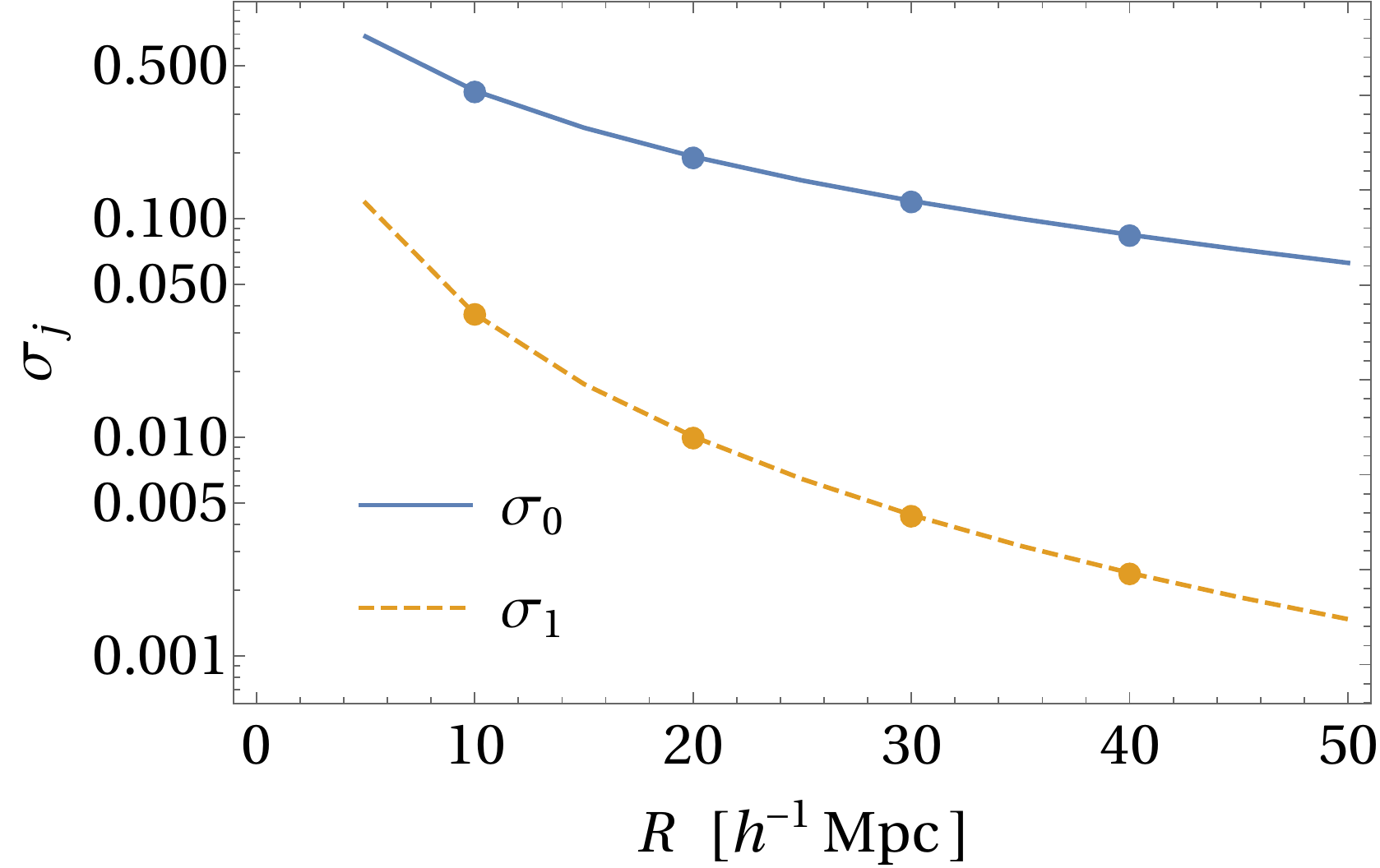}
  \caption{\label{fig:sigma}
    The values of $\sigma_0$ and  $\sigma_1$ calculated from the
    perturbation theory (solid and dashed lines, respectively) and
    simulation data (points with error bars), as functions of
    smoothing radius $R$.
}
\end{figure}
\begin{figure}
  \includegraphics[width=20pc]{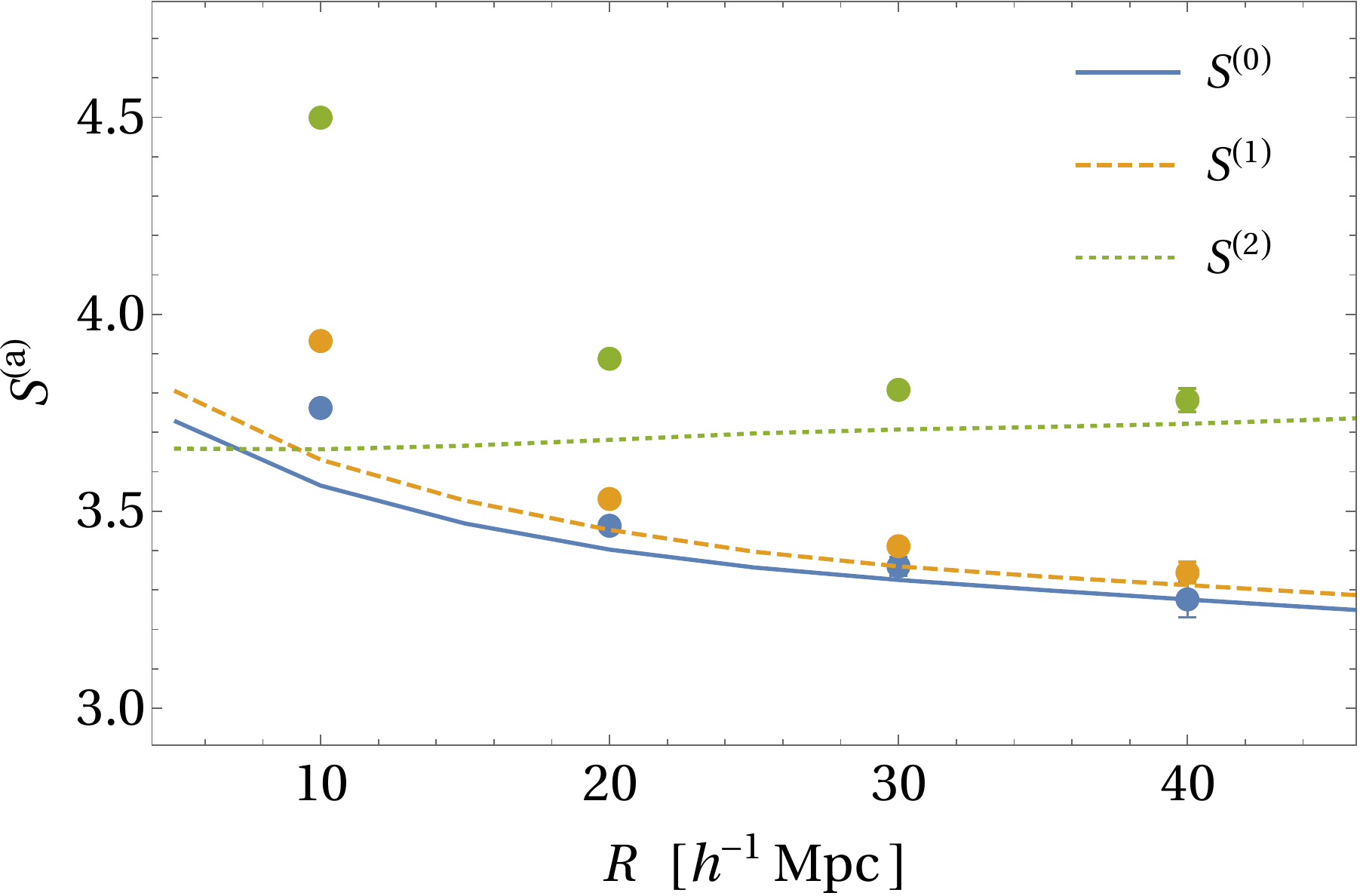}
  \caption{\label{fig:Skew}
    The values of skewness parameters
    $S^{(0)}$, $S^{(1)}$, $S^{(2)}$ calculated from the perturbation
    theory (solid, dashed and dotted lines, respectively) and
    simulation data (points with error bars), as functions of
    smoothing radius $R$.
  }
\end{figure}
\begin{figure}
  \includegraphics[width=20pc]{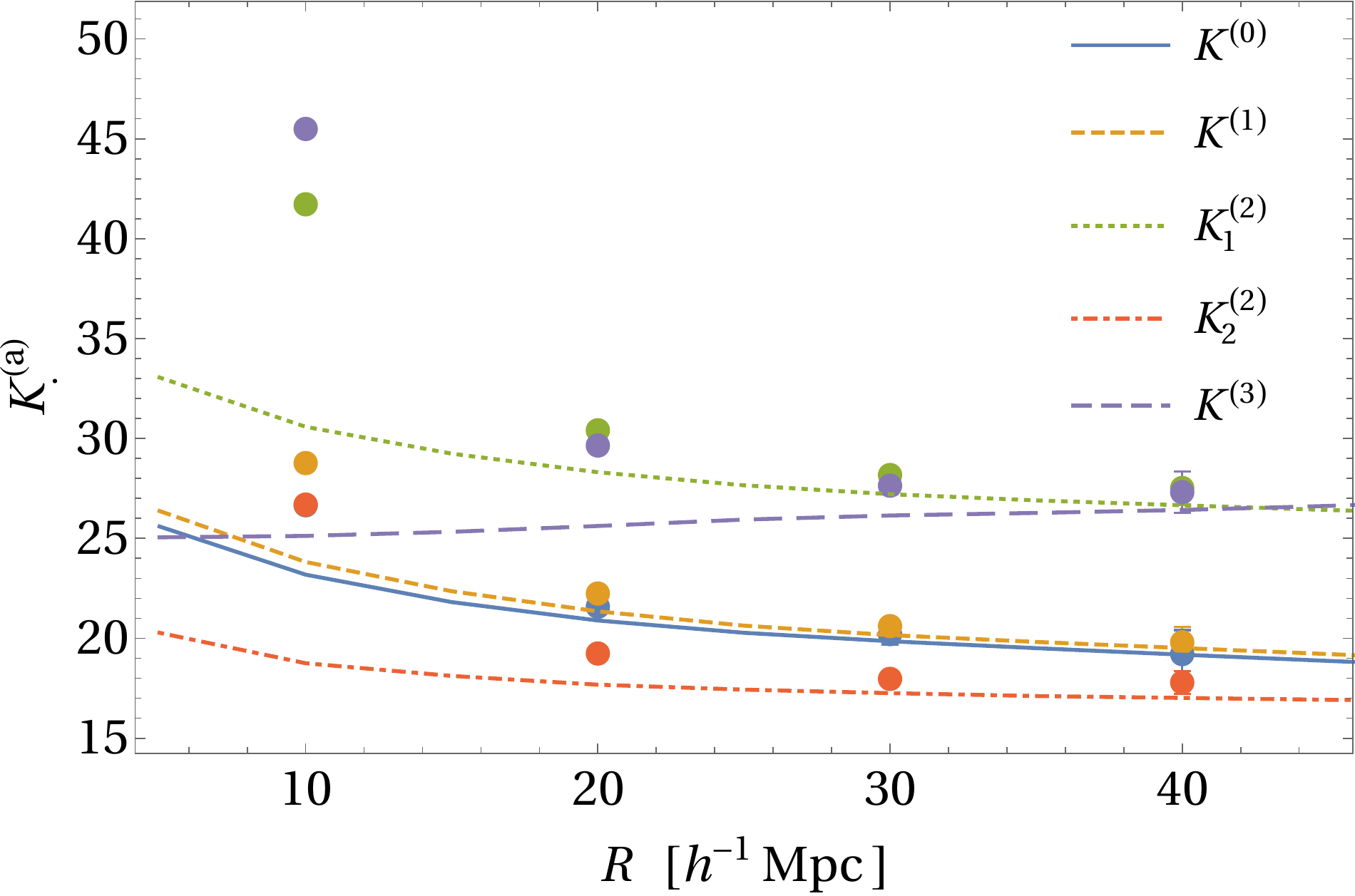}
  \caption{\label{fig:Kurt} The values of kurtosis parameters
    $K^{(0)}$, $K^{(1)}$, $K^{(2)}_1$, $K^{(2)}_2$, $K^{(3)}$
    calculated from the perturbation theory (solid, dashed, dotted,
    dash-dotted and long-dashed lines, respectively) and simulation
    data (points with error bars), as functions of
    smoothing radius $R$. }
\end{figure}

In Table~\ref{tab:params}, the parameters of the weakly non-Gaussian
formula of Minkowski functionals are given. The upper figures of each
entry represent the predictions of the lowest-order perturbation
theory. The lower figures of each entry in Table~\ref{tab:params} are
the values calculated directly from the simulation data with
Eqs.~(\ref{eq:2-4}), (\ref{eq:2-7a})--(\ref{eq:2-9}). In
Figures~\ref{fig:sigma}, \ref{fig:Skew} and \ref{fig:Kurt}, the values
of various parameters calculated from the perturbation theory and
simulation data are compared. The skewness and kurtosis parameters in
the simulations are quantitatively reproduced by the tree-level
perturbation theory on sufficiently large scales within 5\% for
$R\gtrsim 30\,h^{-1}\mathrm{Mpc}$, while the accuracy of the
perturbation theory decreases to 10\% for
$R \sim 20\,h^{-1}\mathrm{Mpc}$ and much worse for
$R\sim 10\,h^{-1}\mathrm{Mpc}$.

\begin{figure*}
  \includegraphics[width=42.5pc]{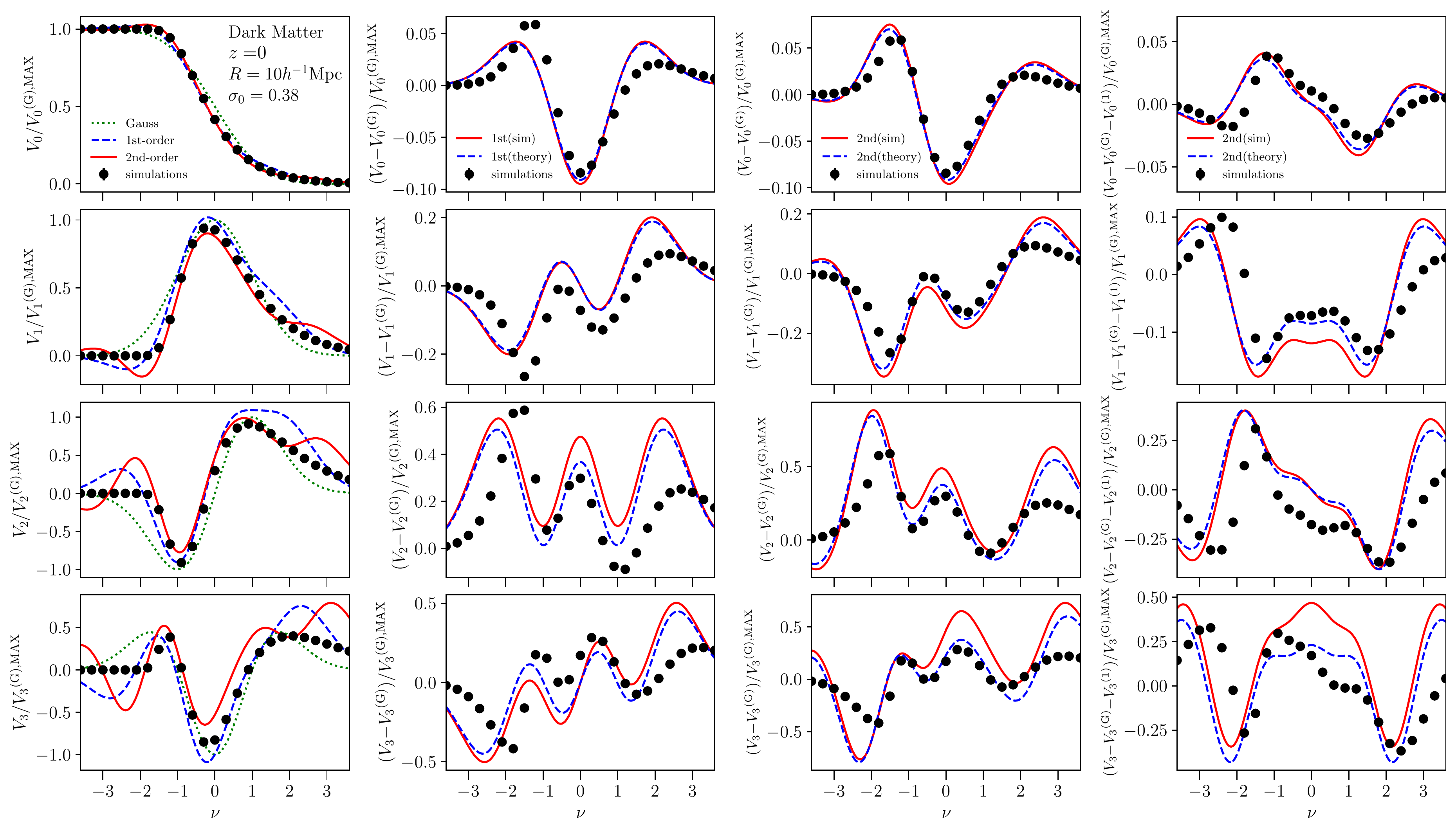}
  \caption{\label{fig:MFs10} Minkowski functionals calculated from the
    numerical simulations (points with error bars) are compared with
    analytic formulas. A smoothing radius $R=10\,h^{-1}\mathrm{Mpc}$
    is adopted. The panels in the leftmost row show the values of
    Minkowski functionals. The curves are normalized by the maximum of
    the absolute values for Gaussian predictions,
    $V_k^\mathrm{(G),MAX}$. The dotted lines correspond to the
    zeroth-order predictions, or Gaussian predictions, dashed lines
    correspond to the first-order predictions, and solid lines
    correspond to the second-order predictions. The parameter values
    for the first- and second-order predictions are numerically
    calculated from the simulations. The panels in the second and
    third rows show differences from the Gaussian predictions. The
    parameter values for the solid lines are taken from the numerical
    simulations, and those for the dashed lines are take from the
    predictions of the perturbation theory. In the panels in the
    second row, the analytic predictions with the first-order
    approximation are presented. In the panels in the third row, the
    analytic predictions with the second-order approximation are
    presented. In the panels in the rightmost row, differences from
    the first-order predictions are plotted.
  }
\end{figure*}
\begin{figure*}
  \includegraphics[width=42.5pc]{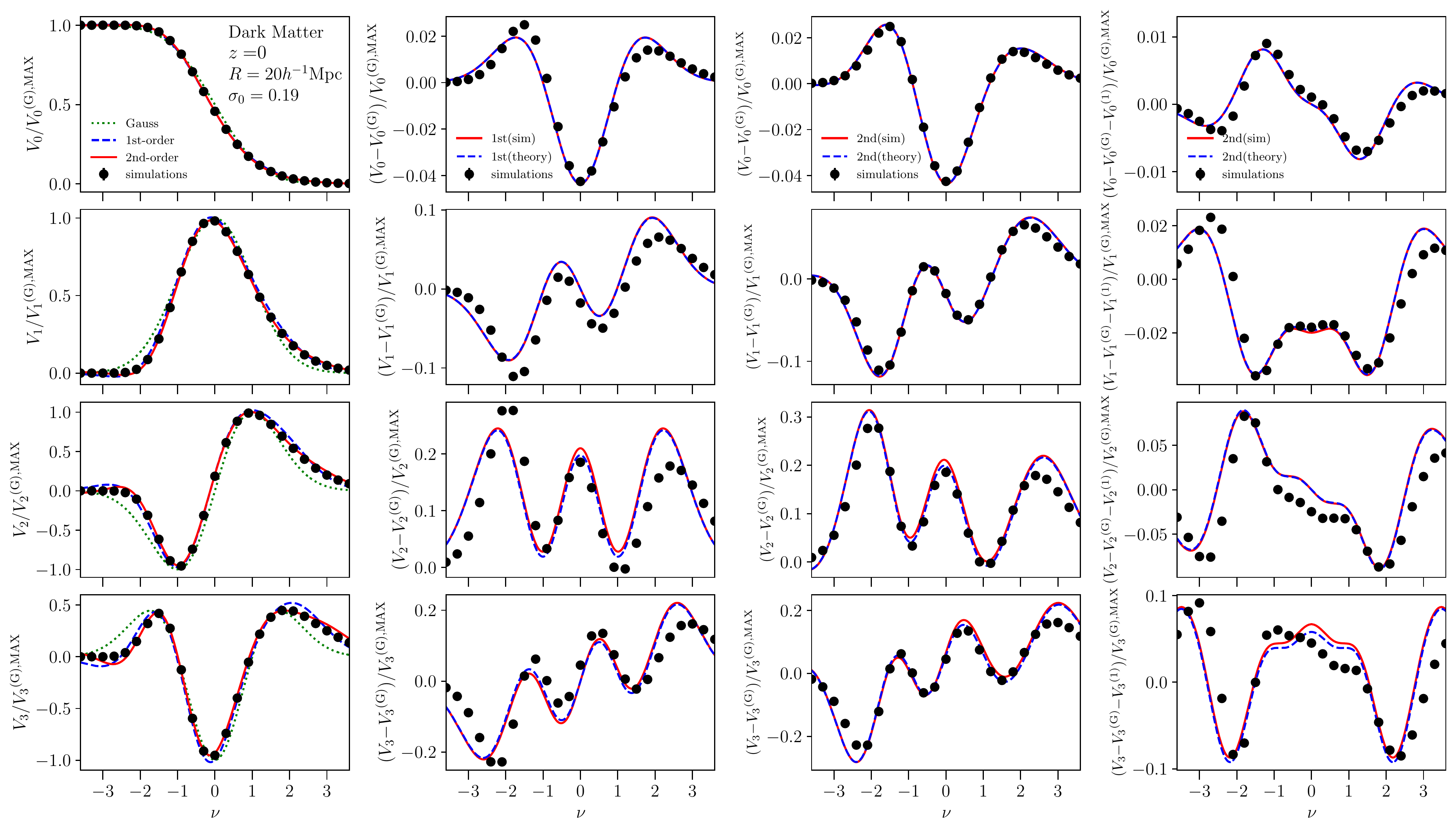}
  \caption{\label{fig:MFs20}
    Same as Figure~\ref{fig:MFs10} but for the smoothing radius of
    $R=20\,h^{-1}\mathrm{Mpc}$.
}
\end{figure*}
\begin{figure*}
  \includegraphics[width=42.5pc]{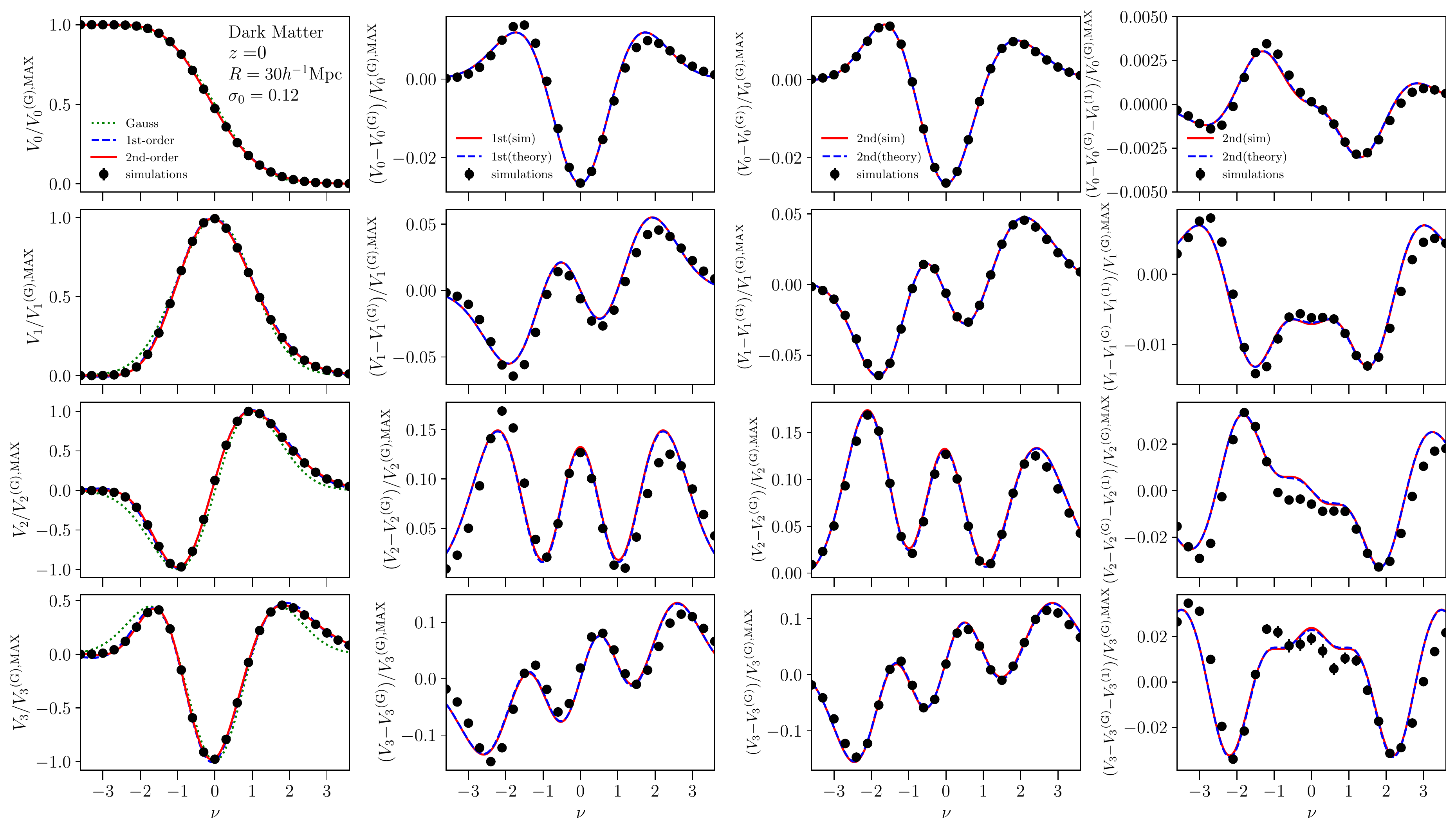}
  \caption{\label{fig:MFs30}
    Same as Figure~\ref{fig:MFs10} but for the smoothing radius of
    $R=30\,h^{-1}\mathrm{Mpc}$.
}
\end{figure*}
\begin{figure*}
  \includegraphics[width=42.5pc]{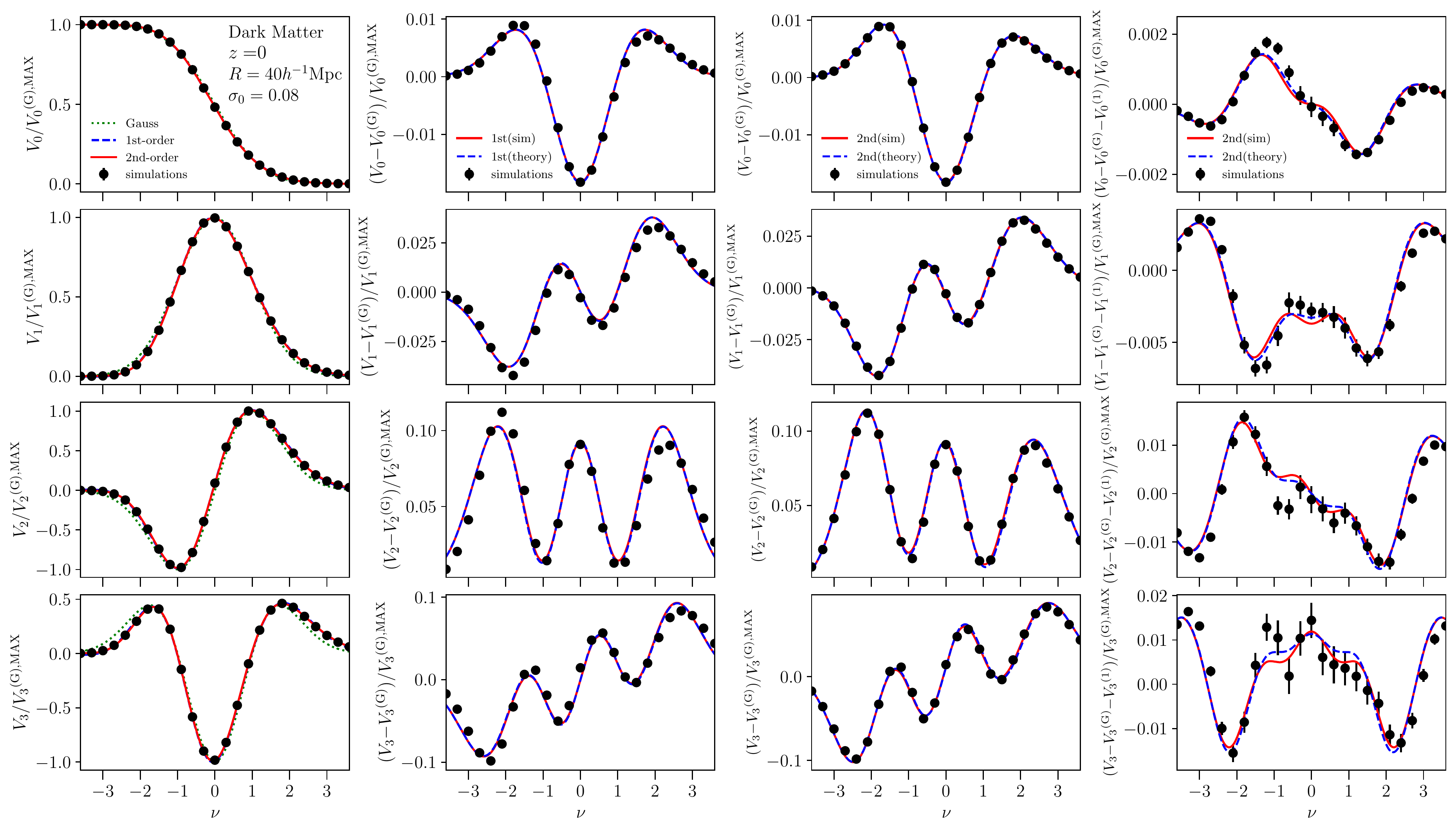}
  \caption{\label{fig:MFs40}
    Same as Figure~\ref{fig:MFs10} but for the smoothing radius of
    $R=40\,h^{-1}\mathrm{Mpc}$.
}
\end{figure*}
Finally, the Minkowski functionals calculated from the numerical
simulations are compared with analytic predictions in
Figures~\ref{fig:MFs10}--\ref{fig:MFs40}. Each figure corresponds to
different smoothing radius. The leftmost rows show the shape of the
Minkowski functionals. The second and third rows show the differences
from the Gaussian predictions, i.e., they depict the non-Gaussian
effects. The lines in the second rows of the figures show the
first-order corrections of non-Gaussianity in the analytic formula.
The lines in the third rows of the figures show the second-order
corrections of non-Gaussianity. The rightmost rows show the differences
from the first-order predictions, i.e., they depict the second- and
higher-order effects of non-Gaussianity in the Minkowski functionals.

As expected, the analytic predictions reproduce the results of
numerical simulations when the smoothing radius is large and the
expansion parameter $\sigma_0$ is small. The qualitative behaviors of
the Minkowski functionals as functions the threshold are reproduced by
the analytic formulas. Quantitatively, however, the agreements are
better for large smoothing radii than for small smoothing radii, as
expected. Comparing the second and third rows of each figure, the
analytic formula with second-order corrections outperforms those with
only first-order corrections. This shows the quantitative usefulness
of taking the second-order effects into account. Purely second-order
effects in the analytic formula are shown in the rightmost rows in the
figures. For the smallest smoothing radius of
$10\,h^{-1}\mathrm{Mpc}$, the performance of the analytic formula with
parameters estimated from the perturbation theory and numerical
simulations are similarly worse than the cases of larger smoothing
radii. This means that the third- and higher-order corrections of the
non-Gaussianity in the analytic formula are not negligible in the case
of smallest smoothing radius.

\section{\label{sec:Conclusions}
  Conclusions
}

In this paper, we compare the second-order formula of weakly
non-Gaussian Minkowski functionals to the results of $N$-body
simulations of the large-scale structure. As expected, the nonlinear
perturbation theory reproduces the deviations from the Gaussian
predictions of Minkowski functionals when the smoothing radius is
large enough. We quantitatively investigate the performance of the
nonlinear perturbation theory against the numerical simulations.

The nonlinear perturbation theory predicts all the parameters in the
analytic formula of weakly non-Gaussian Minkowski functionals. While
the calculations of skewness and kurtosis parameters with the
perturbation theory involve multi-dimensional integrals, parts of the
integrations are analytically performed, and one can numerically
evaluate all the necessary integrals without any difficulty. The
predicted parameters are compared with those directly evaluated by the
$N$-body simulations in Table~\ref{tab:params} and
Figures~\ref{fig:sigma}--\ref{fig:Kurt}.

In our calculations, the nonlinear perturbation theory with tree-level
approximations are adopted. Higher-order corrections of the
perturbation theory with loop corrections may improve the theoretical
predictions, while the numerical evaluations of the multi-dimensional
integrals would be much harder. Investigations along this line is one
of the possible extensions of the present work.

The Figures~\ref{fig:MFs10}--\ref{fig:MFs40} show our comparisons of
the Minkowski functionals between numerical results and analytical
formula for various smoothing radius. As expected, the degree of
agreement varies with smoothing radius. The analytic formula is better
in the larger smoothing radius (i.e., smaller $\sigma_0$), as
expected. While higher-order effects of both non-Gaussianity and the
perturbation theory are simultaneously important for smaller smoothing
radius, the analytic formula with larger smoothing radius outperforms
the case of smaller smoothing radius. 

In this paper, we only consider the clustering of dark matter in real
space, and obviously ignore the effects of galaxy biasing and
redshift-space distortions, which are inevitable in the actual
observations of the large-scale structure of the Universe. While the
purpose of this paper is to investigate the dynamically nonlinear
effects on the Minkowski functionals of density fluctuations of dark
matter, taking into account the biasing and redshift-space distortions
should be necessary to realistically predict the shape of Minkowski
functionals of observable galaxies. We will address these effects in
future work.

Another important application of the present work (with extensions of
including the observational effects mentioned above) is to see whether
or not one can distinguish the primordial non-Gaussianity from the
non-Gaussianity induced by nonlinear evolutions. The method developed
in this paper should offer an analytic way of investigating this kind
of issue in future work.

\begin{acknowledgments}
  This work was supported by JSPS KAKENHI Grants No.~JP19K03835
  (T.M.), No.~JP16K17684 (C.H.), No.~JP16H02792 (S.K.).
\end{acknowledgments}





\twocolumngrid
\renewcommand{\apj}{Astrophys.~J. }
\newcommand{\aap}{Astron.~Astrophys. }
\newcommand{\aj}{Astron.~J. }
\newcommand{\apjl}{Astrophys.~J.~Lett. }
\newcommand{\apjs}{Astrophys.~J.~Suppl.~Ser. }
\newcommand{\apss}{Astrophys.~Space Sci. }
\newcommand{\cqg}{Class.~Quant.~Grav. }
\newcommand{\jcap}{J.~Cosmol.~Astropart.~Phys. }
\newcommand{\mnras}{Mon.~Not.~R.~Astron.~Soc. }
\newcommand{\mpla}{Mod.~Phys.~Lett.~A }
\newcommand{\pasj}{Publ.~Astron.~Soc.~Japan }
\newcommand{\physrep}{Phys.~Rep. }
\newcommand{\ptp}{Progr.~Theor.~Phys. }
\newcommand{\ptep}{Prog.~Theor.~Exp.~Phys. }
\newcommand{\jetp}{JETP }
\newcommand{\jhep}{Journal of High Energy Physics}


\end{document}